\documentstyle[aps,epsf]{revtex}
\begin{document}
\def\btt#1{{\tt$\backslash$#1}}
\draft
\preprint{TPJU -- 25/96}

\title{Deep inelastic scattering of leptons from nuclear targets and the 
       BFKL pomeron}

\author{Andrzej Bialas, Wieslaw Czyz}
\address{Institute of Physics of the Jagellonian University, Reymonta 4, 
         30 - 059 Krak\'ow, Poland}
\author{Wojciech Florkowski}
\address{H. Niewodnicza\'nski Institute of Nuclear Physics, 
         Radzikowskiego 152, 31 - 342 Krak\'ow, Poland}

\date{\today}
\draft

\maketitle

\begin{abstract}
We calculate shadowing in the process of deep inelastic interactions of
leptons with nuclei in the perturbative regime of QCD. We find appreciable
shadowing for heavy nuclei (e.g. Pb) in the region of small Bjorken
scaling variable $10^{-5}\leq x \leq 10^{-3}$.  This shadowing depends
weakly on $Q^2$, but it may be strongly influenced, especially at
$x \geq 10^{-3}$, by the existence of real parts of the forward scattering
amplitudes.
\end{abstract}


\section{Introduction}
\label{sec:intro}

There is considerable progress in calculations of various cross sections 
for deep inelastic lepton scattering in the Lipatov limit ($\nu\to\infty$ 
and $Q^2$ fixed at a large value \cite{FKL75}). In particular a 
picture has been proposed \cite{NZ91,M94} which reproduces the basic features 
of BFKL pomeron, e.g. its Regge behavior which was earlier \cite{FKL75}
obtained from a direct resummation of the Feynman diagrams of Quantum 
Chromodynamics. Subsequently, the model of  \cite{NZ91,M94} was applied 
to calculate the total cross section, the diffractive dissociation 
cross section, and the corresponding structure functions for the proton 
target \cite{NPR95,NPRW96,BP96a,BP96b,B96}.

\medskip
In view of the successful experimental program for the so-called
``small-x physics'' --- which is being pursued at HERA for proton targets
and will be, possibly, extended to nuclear targets --- it makes sense to
adapt the results of \cite{NPR95,NPRW96,BP96a,BP96b,B96} also to
nuclear targets.  So, in this paper, we apply the results of
\cite{NPR95,NPRW96,BP96a,BP96b,B96} to calculate the shadowing in deep
inelastic lepton-nucleus interactions.

\medskip
The problem of nuclear shadowing in deep inelastic lepton-nucleus
interactions has been recently discussed in many papers.  Refs. 
\cite{HERA,NZZ95,PRW95,AB96,ABKSS96,S96} give an up-to-date picture 
of this field.

\medskip
For a description and discussion of the basic physical ideas of the model
used in the present calculations we refer the reader to the article 
\cite{B96}. Here a brief summary will suffice. The model was initially 
designed for ``onium'' - ``onium'' scattering, i.e. for scattering of
two bound $q\bar q$ states with a small transverse separation.  At
high energy --- in perturbative QCD --- such an ``onium'' state evolves
(cascades) into a bunch of colorless dipoles whose number increases as
a power of the incident energy.  The interaction occurs {\it via} two
gluon exchange between the pairs of dipoles (one from each ``onium'').
Navelet, Peschanski, Royon and Wallon \cite{NPR95,NPRW96} applied this
model to deep inelastic lepton-proton scattering.  They assumed ---
naturally --- that the virtual photon at high $Q^2$ can be described by
an ``onium''.  For the target proton --- which certainly does not
resemble an ``onium'' --- they made a rather bold assumption that it can
nevertheless be approximated by a collection of ``onia'' with an
average ``onium'' radius to be determined from the data.  It was thus
rather surprising that such a --- seemingly crude --- model described very
reasonably the existing data on $F_2$ and on gluon structure function
in a fairly large range of $Q^2$ and of Bjorken $x$. To obtain good
fits to the experimental data the authors of Refs. \cite{NPR95,NPRW96} 
had only to normalize $F_2$. We show below that this normalization
corresponds to $n_{eff}\approx 6.7$ ``onia'' in a nucleon.

\medskip
In the present paper we apply these ideas to the calculations of
shadowing in deep inelastic scattering from nuclear targets.  We hope
that this can provide an independent check of the hypothesis of
Ref. \cite{NPR95}, and give more information about the nucleon structure 
at very small Bjorken $x$.

\medskip
The total virtual transverse photon - ``onium'' cross section is one
of the basic inputs in computations of nuclear shadowing and we will
use the following integral representation of
$\sigma^T_{tot}=(\Sigma_nP_n\sigma_n)^T$ ($n$ is the index of the Good
and Walker eigenmodes of absorption, see below)

\begin{equation}
(\Sigma_n P_n\sigma_n)^T = \sigma^T_{tot}(x,Q^2)
= {4N_c\alpha^2n_{eff}\alpha_{em}e^2_f\over\pi}\,r_0^2 
\int\limits^{c+i\infty}_{c-i\infty}{d\gamma\over 2\pi i}\,x^{{-\alpha
N_c\over\pi}\chi(\gamma)} \,2\pi\, 
\left({Qr_0\over2}\right)^{\gamma-2}h^T(\gamma).
\label{eq1}
\end{equation}
This integral representation is our version of $\sigma^T_{tot}$
discussed in \cite{NPR95}. We have checked through numerical
evaluation of (\ref{eq1}) that it agrees well with the results of Refs.
\cite{NPR95,NPRW96} provided $n_{eff}\approx 6.7$.  In (\ref{eq1}) $N_c$ is
the number of colors involved in the process, and $n_{eff}$ is an
effective number of ``onia''in the nucleon.
$\alpha_{em}={1\over137}$, $e^2_f$ is the sum of the squares of quark
charges, and $r_0\approx .8$ fm is approximately the ``onium''
diameter.

\begin{equation}
h^T(\gamma) = 
{4\over\gamma(2-\gamma)} {\Gamma(3-{1 \over
2}\gamma)\Gamma^3(2-{1 \over 2}\gamma)\Gamma(2+{1 \over 2}\gamma)
\Gamma(1+{1 \over 2}\gamma)\over\Gamma(4-\gamma)\Gamma(2+\gamma)}\, ,
\label{eq2}
\end{equation}
and $\chi(\gamma)$ is the eigenvalue of the BFKL kernel defined in
terms of $\psi(\gamma)={d\over d\gamma}[\ln\Gamma(\gamma)]$ as
\begin{equation}
\chi(\gamma)=2\psi(1)-\psi(1-{\textstyle{1 \over 2}}\gamma)-
\psi({\textstyle{1 \over 2}}\gamma)\,\,. 
\label{eq3}
\end{equation}
$\chi(1)$ determines the slope of the BFKL pomeron
\begin{equation}
\Delta_p=\alpha_p-1={\alpha N_c\over\pi}\chi(1)\,\, ,
\label{eq4}
\end{equation}
where $\alpha$ is a strong coupling constant.

\medskip
Since we are interested in the structure function $F_2\sim
(\sigma_{\rm transv}+\sigma_{\rm longitud})$ one should add to
(\ref{eq1}), which gives the transverse virtual photon - ``onium''
total cross section, the contribution of the longitudinal virtual
photon in order to have a complete virtual photon - ``onium'' total
cross section.  As it was shown in Ref. \cite{NPRW96} one gets the
longitudinal photon contribution by replacing the function
$h^T(\gamma)$ of (\ref{eq2}) by $h^L(\gamma)$:
\begin{equation}
h^L(\gamma) = h^T(\gamma){(2-\gamma)\gamma\over
2(2-{\textstyle{1\over 2}}\gamma)
(1+{\textstyle{1\over 2}} \gamma)}\,.
\label{eq5}
\end{equation}
Since the integrand in (\ref{eq1}) is strongly peaked at $\gamma=1$ we can
approximate the above replacement by $h^L(\gamma)\approx
{\textstyle{2\over 9}}h^T(\gamma)$.
Hence the total cross section is 
\begin{equation}
\sigma_{tot}=\sigma^T_{tot}+
\sigma^L_{tot}\approx {\textstyle{11\over9}}\sigma^T_{tot}\,\,.
\label{eq6}
\end{equation}
The success of (\ref{eq1}) (compare \cite{NPR95,NPRW96}) encourages us
to apply the results of \cite{NPR95,NPRW96,BP96a,BP96b,B96} to
evaluation of nuclear shadowing.

\medskip
In the next Section we present the general formula for the shadow in
the deep inelastic virtual photon-nucleus interactions.  In Section
\ref{sec:triple} we apply it to calculate the ``triple pomeron''
\cite{BP96a} contribution to shadowing.  In Section \ref{sec:quasi} 
we do the same for the ``quasielastic'' contributions \cite{BP96b}.  
Section \ref{sec:disc} presents results and conclusions.  Two Appendices 
give details of derivations of some formulae.

\section{Basic formulae for virtual photon-nucleus interactions}
\label{sec:basic}

Many theoretical papers on this and related subjects have been
published in the past and our task is merely to adapt the existing
material for our purposes.  Our main formula, which gives the
deviation from unity of the ratio, $R_A$, of the total virtual
photon-nucleus cross section to the incoherent sum of the cross
sections on individual nucleons, is a slight generalization of the
formula proposed more than 25 years ago by Gottfried and Yennie
\cite{GY69} and then developed by Yennie and Bauer et al
\cite{Y76}. It goes as follows:

\begin{eqnarray}
\Delta_A(x,Q^2) &=& R_A(x,Q^2)-1  \nonumber \\
&=& -{2(A-1)\over\sigma_{tot}} \, \hbox{Re}\, \sum_n P_n
\biggl\{\int d^2b \int\limits_{-\infty}^{+\infty} 
dz_1\int\limits_{-\infty}^{z_1} dz_2 
\left[\, \int d^2b' f_n({\bf b'})\right]^2 \nonumber \\
& &  \times  \, \rho({\bf b},z_1) \, \rho({\bf b},z_2) \, 
e^{i(z_1-z_2)mx_p} 
\left[1-\int\limits_{z_1}^{z_2} dz_3 \, \rho({\bf b},z_3) 
\int d^2b' f_n({\bf b'})\right]^{A-2} \biggr\} \,.
\label{eq7}
\end{eqnarray}
Negative $\Delta_A$ means there is shadowing, positive $\Delta_A$ -
antishadowing.  Here $x={Q^2\over 2m\nu}$ is the Bjorken variable with
$Q^2=-q^2$ (the negative of the four momentum squared of the virtual
photon), $m$ - the nucleon mass, $\nu$ - the energy loss of the
incident lepton (in the rest frame of the target nucleus).
\begin{equation}
x_p={x\over\beta}\,\, ,\,\,\beta={Q^2\over Q^2+M^2}\, ,
\label{eq8}
\end{equation}
where $M$ is the mass of the virtual photon excitation. In this
formula we neglected the correlations between the nucleons inside the
target. The index $n$ stands for the Good and Walker eigenmodes of
absorption \cite{GW60}, which propagate through nuclear matter without
diffractive excitations.  ${\bf b}$ is the impact parameter of the
virtual photon, and $P_n$ is the probability of realization of the
$n$-th eigenmode. $\Sigma_n P_n\sigma_n=\sigma_{tot}$ is the total
transverse virtual photon-nucleon cross section.  $f_n$ is the
forward scattering amplitude of the $n$-th mode.

\medskip
In \cite{BP96a,BP96b,B96} only second order interactions in
dipole-dipole cross section of the diffractively excited systems were
considered.  Then, only two classes of processes seem to be dominant
\cite{B96}: the ``triple pomeron'' contribution where the ``onium'' 
scatters inelastically, and the ``quasielastic''contribution where
the ``onium'' scatters quasielastically from the target proton.

\medskip
Note that the results of Refs. \cite{BP96a,BP96b,B96} we are going to
use to calculate (\ref{eq7}) do not give us $P_n$ and $f_n({\bf b})$
but only $\Sigma_n P_n f_n({\bf b})f_n({\bf b})$. However, $\Sigma_n
P_n f_n({\bf b})f_n({\bf b})f_n({\bf b})$ and higher order expressions
have, so far, not been calculated (e.g. in the case of an exchange of
three pomerons one would have to calculate a three dipoles distribution
function), hence the last factor in (\ref{eq7}), $[...]^{A-2}$, can,
at present, be only approximated (see below).

\medskip
Note also that when the transverse spatial extension of $f_n$ is much
smaller than the nuclear radius (hence the transverse extension of
$\rho({\bf b},z)$) we can approximate 

\begin{equation}
f_n({\bf b})=\delta^{(2)}({\bf b})\int d^2b^{\prime} 
f_n({\bf b^{\prime}})\,. 
\label{eq9}
\end{equation}
This approximation was employed in (\ref{eq7}).  Physically, this is
the question of the relative transverse sizes of the cascade of dipoles
\cite{BP96a,BP96b,B96} and of the nucleus.  We have checked accuracy
of the assumption (\ref{eq9}) for $\rho({\bf b},z)$ suitable for large
nuclei (see Section \ref{sec:disc}) and found it acceptable; 
$\rho({\bf b},z)$'s are taken in form of Saxon-Woods distributions.

\medskip
Now we will discuss the other input, $\Sigma_nP_nf_nf_n$ , from 
\cite{BP96a,BP96b,B96} and to make it concise we shall proceed through 
numerous references to many formulae of the papers of Refs.
\cite{BP96a,BP96b,B96}.

\section{``Triple pomeron'' contribution to the shadowing}
\label{sec:triple}
   
We assume $A\gg1$, hence a large nucleus where (\ref{eq9}) is
applicable. We first work out only the double scattering factor in
(\ref{eq7}).  We modify the formula (21) of \cite{BP96a} introducing
two impact parameters ${\bf\tilde b}$,${\bf\tilde b'}$ (to describe
collisions with two different nucleons) and taking the inverse Mellin
transformation of both sides of (21):

\begin{eqnarray}
& &\int d^2\tilde b d^2\tilde b'(\Sigma'_nP_nf_n({\bf\tilde b})
f_n({\bf\tilde b'})) \nonumber \\
&=& 2\pi \, \alpha^5 n^2_{eff} N_c \, x^{-2\Delta_p}_p 
\left({2a_p\over \pi}\right)^3 
\int\limits^{c+i\infty}_{c-i\infty} {d\gamma\over2\pi i}
\, \rho^{2-\gamma} \beta^{-\alpha N_c\chi(\gamma)/\pi} \nonumber \\
& & \times \!\! \int {d^2\tilde b \, d^2\tilde b' \over
\tilde b^2\tilde b'^2}
\int dx_{12}dx_{02} W(x_{12},x_{02})D({\bf \tilde b},x_{02})
D({\bf \tilde b'},x_{12})\,,
\label{eq10}
\end{eqnarray}
where
\begin{equation}
a_p=[7\alpha N_c\zeta(3)\ln(1/x_p)/\pi]^{-1}=a(x_p)\, ,
\label{eq11}
\end{equation}
and $\Sigma'_n$ denotes a partial summation over the parameters
specifying diffractive eigenstates (for details see \cite{BP96a}). The
prime is added because the summation over the masses, $M$, of the
diffractively excited states is left till the very
end. $W(x_{12},x_{02})$ is a symmetric function of $x_{12}$ and
$x_{02}$, and e.g. for $x_{02}<x_{12}$
\begin{equation}
W(x_{12},x_{02})=x_{12}^{\gamma-2}F\left(1-{\textstyle{1\over 2}}\gamma,
1-{\textstyle{1\over 2}}\gamma;1;\left({x_{02}\over x_{12}}\right)^2\right) 
\,\, , \label{eq12}
\end{equation}
$F$ being the hypergeometric function, and
\begin{equation}
D(\tilde b,x)\!=\!\!\int\!\! d^2rdz\Phi(r,z)r\ln
\left({\tilde b^2 \over rx}\right)
\exp\left[-{a_p\over 2} \ln^2\left({\tilde b^2 \over rx}\right)\right],
\label{eq13}
\end{equation}
where $\Phi(r,z)$ is the square of the proton wave function.  For
small $a_p$ one can evaluate (\ref{eq10}) analytically (the relevant
integrals are given in Appendix A) and obtain for the transverse
virtual photon

\begin{eqnarray}
& & \int d^2\tilde b d^2\tilde b'\Sigma'_nP_nf_n({\bf\tilde b})
f_n({\bf\tilde b'}) \nonumber \\
& & \equiv \langle f^2_n \rangle (\rho,\beta,x_p) = 
32\pi \, \alpha^5 n^2_{eff} N_c \, x_p^{-2\Delta_p}
\left({a_p\over\pi}\right)^{1\over 2} \int\limits_{c-i\infty}^{c+i\infty}
{d\gamma\over2\pi i}\rho^{2-\gamma}
\beta^{-\alpha N_c\chi(\gamma)/\pi}V(\gamma)r_0^{2+\gamma}
e^{\gamma^2\over4a_p} \,\, .
\label{eq14}
\end{eqnarray}
where
\begin{equation}
V(\gamma)=\int\limits^1_0 F(1-{\textstyle{1\over 2}}\gamma,
1-{\textstyle{1\over 2}}\gamma;1;y^2)dy\,\,,
\label{eq15}
\end{equation}
and $r_0=\int d^2rdz\Phi(r,z)r $ is a non-perturbative parameter
determined from the data. A fit to the proton structure function gives
$r_0\approx .8$ fm \cite{NPR95,NPRW96}.

\medskip
At this point we introduce the nuclear algorithm with $A$ collisions
inside the nucleus. We write 
\begin{equation}
t_A = \langle f_n^2 \rangle \biggl\{1-f_{eff}
\int\limits^{z_2}_{z_1} dz_3 \,
\rho({\bf b},z_3)\biggr\}^{A-2}\,\, ,
\label{eq16}
\end{equation}
and
\begin{equation}
T_A = A(A-1)\int d^2b\int\limits_{-\infty}^{+\infty}dz_1
\int\limits_{-\infty}^{z_1}dz_2
e^{i(z_1-z_2)mx_p}
\, \rho({\bf b},z_1) \, \rho({\bf b},z_2) \, t_A\,.
\label{eq17}
\end{equation}

The question now is what to take for $f_{eff}$ ?  As we have already
remarked the results given in \cite{BP96a,BP96b,B96} do not go beyond
the double scattering.  Also, \cite{BP96a,BP96b,B96} do not provide us
with amplitudes for the eigenstates of absorption but with amplitudes
averaged over many eigenstates.  In order to have such amplitudes one
would have to do new and --- possibly --- rather complicated calculations.
Therefore, in order to {\it estimate} the multiple scattering
corrections we shall use the full amplitude for dipole-proton
scattering at fixed $\rho$ (this is the diameter of the ``onium''
representing the virtual photon)

\begin{equation}
f_{eff}(r_0,\rho,x_p)=\int d^2\tilde b \, T(r_0,\rho,\tilde b,x_p)\,\, ,
\label{eq18}
\end{equation}
where \cite{BP96b}
\begin{equation}
T(r_0,\rho,\tilde b,x_p) = 
\pi\alpha^2n_{eff}{r_0\rho\over\tilde b^2}
\ln\Bigl({\tilde b^2\over r_0\rho}\Bigr)x_p^{-\Delta_p}
\Bigl({2a_p\over\pi}\Bigr)^{3\over 2}
e^{-{\textstyle{1\over 2}} a_p\ln^2({\tilde b^2\over r_0\rho})}\,\,.
\label{eq19}
\end{equation}
The integration over $d^2\tilde b$ can be done (see Appendix A) and we
obtain 
\begin{equation}
f_{eff}(r_0,\rho,x_p) = 
2^{3/2}\pi\alpha^2n_{eff} r_0\rho x_p^{-\Delta_p}
\Bigl({a_p\over\pi}\Bigr)^{{\textstyle{1\over 2}}}
e^{-{\textstyle{1\over 2}} a_p\ln^2({r_0\over\rho})}\,\,.
\label{eq20}
\end{equation}

A comment is here in order. To estimate the multiple scattering
corrections one can use some other than $f_{eff}$ effective
amplitude, e.g. the one proposed in Ref. \cite{S96} Eq. (5).
In our notation it amounts to taking the ratio $\langle f_n^2 \rangle
/ f_{eff}$ instead of $f_{eff}$. As expected, $\langle f_n^2 \rangle
/ f_{eff} > f_{eff}$ but both are of the same order of magnitude.
Since the complete expressions for multiple scattering corrections
need direct calculations of the contributions of interactions with
3 and more nucleons which are not done, we choose $f_{eff}$ as a
somewhat simpler expression.

The last point is to average $T_A$ over the {\it probability},
$\tilde\Phi$, to find a dipole in the virtual transverse or
longitudinal photon \cite{NZ91}:
\begin{mathletters}
\label{eq21}

\begin{equation}
\tilde\Phi^T(\hat z,\rho,Q) = 2{N_c\alpha_{em}\over(2\pi)^2}e_f^2
(\hat z^2+(1-\hat z)^2)\hat Q^2 K_1^2(\hat Q\rho)\,\,, 
\label{eq21a}
\end{equation}

\begin{equation}
\tilde\Phi^L(\hat z,\rho,Q) = 8 {N_c\alpha_{em}\over(2\pi)^2}e^2_f
\hat z(1-\hat z)\hat Q^2 K^2_0(\hat Q\rho)\,\, .
\label{eq21b}
\end{equation}

\end{mathletters}
\noindent Here
\begin{equation}
\hat Q=\sqrt{\hat z(1-\hat z)}Q\;\;\;.
\label{eq22}
\end{equation}

Thus the ``triple pomeron''(TP) transverse (T) and longitudinal (L)
contributions to the shadow, within the approximation (\ref{eq9}), is
as follows

\begin{eqnarray}
& & \Delta^{TP(T,L)}_A(x,Q^2) \nonumber \\
& & = -{2(A-1)\over\sigma_{tot}}\int_{10x}^1 
d\beta \,\hbox{Re}\, \biggl\{\int_0^1 d\hat z\int_0^{r_0} d^2\rho 
\,\tilde\Phi^{T,L}(\hat z,\rho) \int d^2b\int^{+\infty}_{-\infty}
dz_1\int^{z_1}_{-\infty}dz_2 \nonumber \\
& & \times \,\, e^{i(z_1-z_2)mx_p} \rho({\bf b},z_1) \rho({\bf b},z_2) 
\langle f^2_n \rangle 
(\rho,\beta,x_p) 
\left[1-f_{eff}(r_0,\rho,x_p)
\int^{z_2}_{z_1}dz_3\rho({\bf b},z_3)\right]^{A-2}\biggr\}   \,\,\, .
\label{eq23}
\end{eqnarray}
For  comments on the limits of the integration over $\beta$ and $\rho$ see
Section \ref{sec:disc}.

\medskip
It is convenient to extract from (\ref{eq23}) a correction factor,
$C_{ms}(x,Q^2)$, for multiple interactions higher than double.  This
is the factor which multiplies the double scattering contribution to
give $\Delta^{TP}_A(x,Q^2)$. Thus we have

\begin{eqnarray}
& & \Delta^{TP(T,L)}_A(x,Q^2) \nonumber \\
& & = -C_{ms}(x,Q^2){2(A-1)\over\sigma_{tot}} \int_{10x}^1 
d\beta \,\hbox{Re}\, \biggl\{\int_0^1 d\hat z\int_0^{r_0} d^2\rho 
\tilde\Phi^{T,L}(\hat z,\rho) \int d^2b\int^{+\infty}_{-\infty}
dz_1\int^{z_1}_{-\infty}dz_2 \nonumber \\
& & \times \,\, e^{i(z_1-z_2)mx_p} \rho({\bf b},z_1)\rho({\bf b},z_2) 
\langle f^2_n \rangle 
(\rho,\beta,x_p)\,\,.
\label{eq24}
\end{eqnarray}
$C_{ms}(x,Q^2)$ will be used in the next section to compute the shadow
of the whole multiple scattering process from the shadow of the double
scattering contribution.  We believe that the level of accuracy of our
calculations and the size of $C_{ms}(x,Q^2)$ justify this approximation.

\section{The quasielastic contributions to the shadowing}
\label{sec:quasi}

This time the Ref. \cite{BP96b} provides us with the basic formulae
and we will assume $A\gg 1$ hence, again, we accept the approximation
(\ref{eq9}).  The method used in \cite{BP96b} to obtain the double
scattering contribution is different than in the case of ``triple
pomeron'' \cite{BP96a} and we now have at our disposal the probability
{\it amplitude} for the intermediate state which, in the present case,
is the one dipole state.  Therefore, the procedure of integration over
$\tilde b$ is done at the level of the amplitude which, subsequently,
is contracted with the probability amplitude for finding a dipole in a
photon of a given polarization, and the result squared.

So, starting from the formulae (19) and (20) of Ref. \cite{BP96b}, we
first replace ${d\sigma_{T,L}\over dM^2}$ by ${d\sigma_{T,L}\over
d\beta}={Q^2\over\beta^2}{d\sigma_{T,L}\over dM^2}$, do the
integration of the amplitude $T$ over $\tilde b$, and obtain the
following double scattering terms in the approximation (\ref{eq9}):

\begin{mathletters}
\label{eq25}
\begin{eqnarray}
& & {d\tilde\sigma_T\over d\beta} = {N_c\alpha_{em}e^2_f\over2\pi}
{Q^4\over\beta^2}\int^1_0 d\hat z\hat z^2(1-\hat z)^2 
[\hat z^2+(1-\hat z)^2] 
\left[ \int^{r_0}_0 d\rho\rho f_{eff}(r_0,\rho,x_p)
K_1(\hat Q\rho)J_1(\hat M\rho)\right]^2 \nonumber \\
\label{eq25a}
\end{eqnarray}
and
\begin{equation}
{d\tilde\sigma_L\over d\beta} = 4{N_c\alpha_{em}e^2_f\over2\pi}
{Q^4\over\beta^2}\int^1_0 d\hat z\hat z^3(1-\hat z)^3 
\left[ \int^{r_0}_0 d\rho\rho f_{eff}(r_0,\rho,x_p)
K_0(\hat Q\rho)J_0(\hat M\rho)\right]^2\,\,, \label{eq25b}
\end{equation}
\end{mathletters}
where $f_{eff}(r_0,\rho,x_p)$ is given by (\ref{eq18}), and $\hat
M=\sqrt{\hat z(1-\hat z)}M$, with $M$ being the mass of the virtual
photon excitation.

\medskip
In the double scattering cross sections (\ref{eq25}) $f_{eff}$ appears
quadratically.  Therefore, since the basic ingredient of
$C_{ms}(x,Q^2)$ defined in Section \ref{sec:triple} is the factor
\mbox{$[1-f_{eff}\int^{z_2}_{z_1}dz_3 \rho({\bf b},z_3)]^{A-2}$} (defined
through the same amplitude $f_{eff}$ as in (\ref{eq25})), we accept
that by multiplying $\Delta^{QE}_A$ for just the double interaction
inside the target nucleus by $C_{ms}$ we obtain a reasonable
approximation for $\Delta^{QE}_A$ with all possible (i.e. $A$)
interactions included.  The error made in this approximate procedure
is not significant because $C_{ms}$ is close to unity throughout the
relevant region of $x$ and $Q^2$ (see Fig. 3).

\medskip
Following steps similar to those which led us to the formula
(\ref{eq23}) we obtain for the quasielastic contributions to the
shadow of the transverse and longitudinal virtual photons the following
expressions:

\begin{mathletters}
\label{eq26}
\begin{eqnarray}
& & \Delta^{QE(T)}_A(x,Q^2) \nonumber \\
& & = -{2(A-1)\over\sigma_{tot}} \,\,
C_{ms}(x,Q^2)
\int^1_{10x} \! d\beta \,\hbox{Re} \biggl\{\int d^2b\int^{+\infty}_{-\infty} 
\!\!\! dz_1\int^{z_1}_{-\infty} \!\!
dz_2 e^{i(z_1-z_2)mx_p} \rho({\bf b},z_1) \, \rho({\bf b},z_2) \nonumber \\
& & \times \, {N_c\alpha_{em}e^2_f\over2\pi}
{Q^4\over\beta^2} \int^1_0 d\hat z\hat z^2(1-\hat z)^2
[\hat z^2+(1-\hat z)^2] 
\left[\int^{r_0}_0d\rho\rho f_{eff}(r_0,\rho,x_p)
K_1(\hat Q\rho)J_1(\hat M\rho)\right]^2\biggr\} \nonumber \\
\label{eq26a}
\end{eqnarray}
and
\begin{eqnarray}
& & \Delta^{QE(L)}_A(x,Q^2) \nonumber \\
& & = -{2(A-1)\over\sigma_{tot}} \,\,
C_{ms}(x,Q^2)
\int^1_{10x} \! d\beta \,\hbox{Re}\, 
\biggl\{\int d^2b\int^{+\infty}_{-\infty}
\!\!\! dz_1\int^{z_1}_{-\infty} \!\!
dz_2 e^{i(z_1-z_2)mx_p} \rho({\bf b},z_1) \rho({\bf b},z_2) \nonumber \\
& & \times {4N_c\alpha_{em}e^2_f\over2\pi}
{Q^4\over\beta^2} \int^1_0 d\hat z\hat z^3(1-\hat z)^3 
\left[\int^{r_0}_0d\rho\rho f_{eff}(r_0,\rho,x_p)K_0(\hat Q\rho)
J_0(\hat M\rho)\right]^2\biggr\}\,, \nonumber \\
\label{eq26b}
\end{eqnarray}
\end{mathletters}
where $f_{eff}(r_0,\rho,x_p)$ is given by (\ref{eq20}).

\medskip
Clearly, setting $C_{ms}=1$ reduces both (\ref{eq24}) and (\ref{eq26})
to the correct expressions for the double scattering contributions to
the shadow.  In fact --- as we have already stressed --- the double
scattering term is more reliable than the multiple scattering
correction because its input from \cite{BP96a,BP96b,B96} comes
exclusively from the ``first principles''.  That is why it is of
interest to discuss deuterium target. This will be discussed in a
forthcoming paper.

\section{Discussion of the results and the conclusions}
\label{sec:disc}

Before presenting any results a comment on the real parts of the amplitudes
$f$ is in order.  First, the dependence of $\sigma_{tot}$ on $x$ , hence
on $2m\nu\approx s$ implies that $\hbox{Re}[f]$ must be different from zero.
This follows from the derivative analyticity relations for the forward
elastic hadronic amplitudes worked out in e.g. Ref. \cite{SS75}.
We can introduce, approximately, this effect into all our expressions
derived so far performing the following replacements:

\begin{eqnarray}
\langle f^2_n \rangle  &\longrightarrow& \langle f^2_n \rangle
(1-i\hat\alpha)^2, \nonumber \\
f_{eff} &\longrightarrow&  f_{eff} (1-i\hat\alpha),
\label{eq27}
\end{eqnarray}
where $\hat\alpha$ is the ratio of the real to imaginary parts of the
forward amplitude.  We use the leading term of the expansion of the
forward amplitudes $f$ given in \cite{SS75}:

\begin{equation}
\hat\alpha={\hbox{Re} f\over\sigma_{tot}}\approx 
\tan\left({\textstyle{1 \over 2}} \pi\Delta_p\right)\,\,.
\label{eq28}
\end{equation}
We accept $\Delta_p\approx .3$, hence $\hat\alpha\approx .5$ .

\medskip
Now the results. We have calculated all contributions to the shadow
$\Delta_A$ from the triple pomeron and the quasielastic diffractions
for two target nuclei: Lead (Pb,A=208) and Calcium (Ca,A=40).  All
integrations in (\ref{eq23}) and (\ref{eq26}) were done numerically
with an extensive help of {\sl Mathematica} (more details are given in
Appendix B).

\medskip   
The single nucleon densities $\rho({\bf s},z)$ were taken in the standard
form
\begin{equation}
\rho({\bf s},z)={1\over V(1+e^{(r-r_A)/a})} \,\,\, ,
\label{eq29}
\end{equation}
with $r=\sqrt{{\bf s}^2+z^2}$, $r_A=1.2 \, \hbox{fm} \, A^{1/3}$, and 
$a= .54 \, \hbox{fm}$.
\begin{equation}
V= \int d^2sdz(1+e^{(r-r_A)/a})^{-1}\,\, ,
\label{eq30}
\end{equation}
so that $\int d^2sdz \rho({\bf s},z)=1$.

\medskip   
We assumed $\Delta_p= .3$ which , through relation (\ref{eq4}) and using
$N_c=3$ gives the strong coupling constant $\alpha= .11.$  This is an
acceptable set of parameters characterizing the BFKL pomeron.  We set
$n_{eff} = 6.7$ to recover the fit to the experimental data of the
$F_2$ of the nucleon \cite{BP96a,BP96b,B96}.

\medskip
In all integrations we restricted the values of $x_p$ to

\begin{mathletters}
\label{31}
\begin{equation}
x_p={x\over\beta}\leq .1 \,\,\,,
\label{eq31a}
\end{equation}
which, as seen from (\ref{eq7}), amounts to excluding distances $(z_1
- z_2)\leq 2 \, {\rm fm}$ from the double- and higher multiplicity
interactions; physically a reasonable condition to impose.  This was
done because in construction of the triple pomeron contribution to
$\Delta_A$, Eq. (\ref{eq23}), see also Appendix A, we explicitly used
the approximation of small $a_p$, hence small $x_p$.  We have checked
that changing the upper limit for $x_p$ (\ref{eq31a}) by a factor
$1.5$ or $2$ does not change significantly $\Delta^{TP}_A$, at the
values of $x$ for which we have done the calculations, namely $x \leq
.01$.  This gives the lower limit for integration over $\beta$:
$\beta>10x$.  

\medskip
The upper limit for $\beta$ on the other hand, should be
taken close to unity because small masses give important contributions
to shadowing. So, the integration range for $\beta$ is taken as
\begin{equation}
10x < \beta < 1\,\,\,. 
\label{eq31b}
\end{equation} 
\end{mathletters}

\medskip
All this makes almost all {\it asymptotic} formulae for the ``triple
pomeron'' given in Refs. \cite{BP96a,BP96b,B96} inacceptable for our
purposes because they lead to infinities for integrations over
$\beta$'s close to unity .

\medskip
We wish to make it clear: by extending our calculation to $\beta$
close to $1$, hence to very small masses $M$, we go beyond the region
of $\beta$'s for which the original formulation of Refs.
\cite{NPR95,NPRW96,BP96a,BP96b,B96} was devised.  We make this extrapolation 
because {\it the complete integral representations} {\it of various} 
$\Sigma_nP_nf_nf_n${\it given also in} \cite{BP96a,BP96b,B96} {\it lead to 
well} {\it defined finite results, even for} $\beta\to 1$ and only they 
are employed in our calculations.  The price to pay is loss of simple
analytic expressions and necessity of numerical integrations.
Although the asymptotic form of $\sigma_{tot}=\Sigma_nP_n\sigma_n$
does not lead to any such problems, to be consistent with our
treatment of $\Sigma_nP_nf_nf_n$ , we also used its integral
representation (\ref{eq1}).

\medskip
We set the cutoff parameter, $r^*$, of the integration over $\rho$,
following the discussion of Ref. \cite{BP96b}, at $r^* = r_0$.
Unfortunately, in the case of nuclear targets the saturation of the
values of the integrals over $\rho$ does not look as good as in the
case of the nucleon target (see Ref. \cite{BP96b}, Fig. 2).
Therefore, we should treat $r^*$ as a parameter with the lower bound
equal $r_0$.  This implies that taking $r^*= r_0$ we tend to
underestimate the shadow.

\medskip
One more general comment: we accept that the contribution to
$\Delta_A$ of the triple pomeron - , of the quasielastic transverse -
, and of the quasielastic longitudinal processes are additive.  In
other words we do not introduce any interferences between them.

\bigskip
In Fig. 1 a,b the total shadow
$\Delta_A=\Delta^{TP(T)}_A+\Delta^{TP(L)}_A+\Delta^{QET}_A+\Delta^{QEL}_A$
is presented for Pb and Ca targets.  As we can see, the pattern of all
curves for these two targets is very similar, only the size of the
shadow for Pb is a factor of two larger than for Ca. We have also
checked through explicit calculations that all components of
$\Delta_A$ of Pb are by the same factor $(\approx \,2)$ larger than of
Ca. Therefore, we will, henceforth, show only the results for Pb.

\medskip
In Fig. 2 a,b we show $\Delta^{TP}_A$ with (a) and without (b)
presence of the real part of the forward scattering
amplitudes. Comparing this figure with Fig. 1 we see that
$\Delta^{TP}_A$ dominates $\Delta_A$, and that the presence of $\hat
\alpha\not=0$ makes shadowing smaller.  In fact a different from zero
real part makes $|\Delta_A|$ smaller: by 16\% at very small $x$ and by
60\% (!) at $x= .01$ . (The  role of the real parts of the forward
amplitudes at various $x$ has been recently discussed in \cite{S96,BC94}.)

\medskip
In Fig. 3 a,b we show the correction factor for multiple scattering
higher than double, $C_{ms}(x,Q^2)$, with (a) and without (b) presence
of $\hat\alpha$.  We can see that in either case $C_{ms}$ is close
enough to unity to justify our procedures in construction of
$\Delta_A$.

\begin{figure}[htbp]
\bigskip
\centerline{\epsfxsize=13.0cm \epsfbox{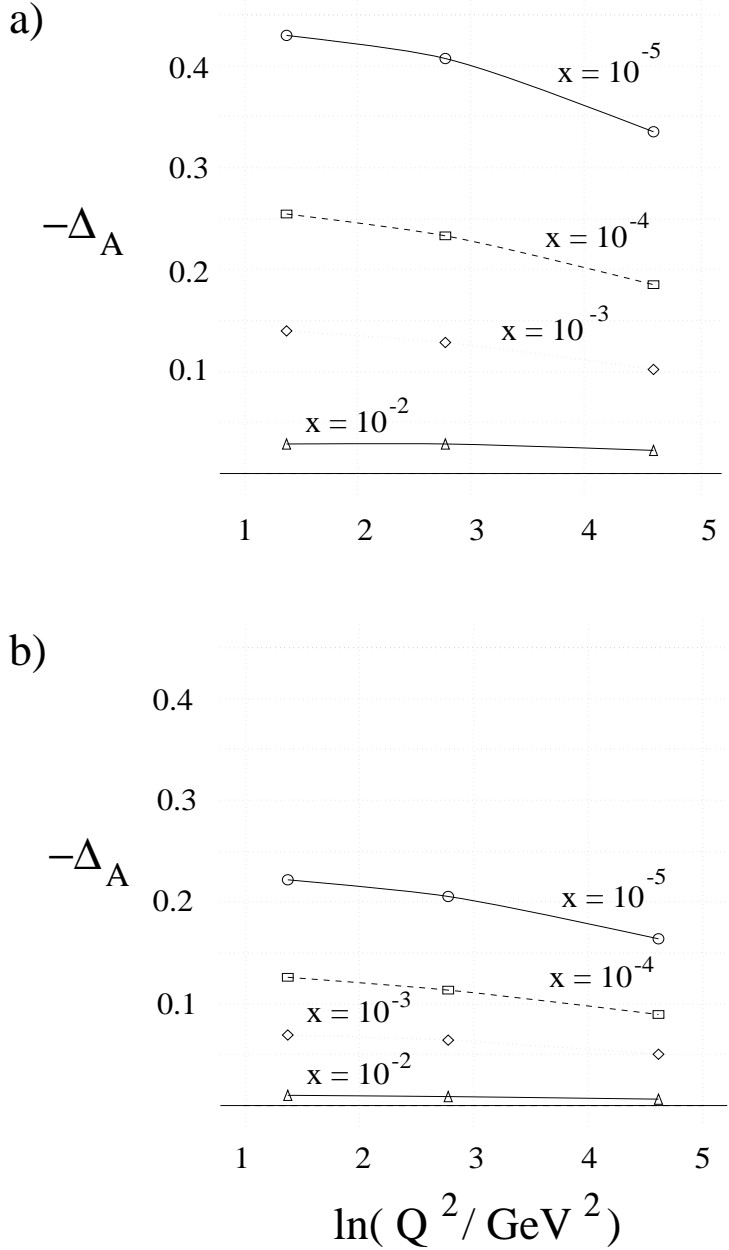}}
\bigskip
\caption{ The total
shadow, $\Delta_A$, for (a) Pb and (b) Ca nuclei. $-\Delta_A$ is
plotted {\it vs} $\ln(Q^2/\hbox{GeV}^2)$ for four values of the
Bjorken scaling variable $x$.  The ratio of the real to imaginary
parts of the forward amplitudes is set at
$\hat\alpha=\tan({\textstyle{1\over 2}}\pi\Delta_p)\approx .5$ .}
\end{figure}

\medskip
The following conclusions result from inspection of the Figs. 1 --- 3:

(i) In the region $x <10^{-3}$ our calculation shows an appreciable 
shadowing, increasing with decreasing $x$. The $x$ dependence of $\Delta_A$
follows approximately the power law

\begin{equation}
\Delta_A \,\, \sim  \,\, x^{-\lambda},
\label{eq32}
\end{equation}
with $\lambda \approx 0.25$, somewhat smaller than the assumed value of
$\Delta_p = 0.3$. An indication of saturation is seen at 
$Q^2 = 4 \, \hbox{GeV }^2 \,$.

(ii) $\Delta_A$ decreases slowly with increasing $Q^2$. The effect is somewhat
stronger at small $x$ and large $Q^2$.

(iii) Multiple scattering corrections account for less than 20 percent
of the value of $\Delta_A$.

(iv) The existence of the real parts of the forward amplitudes makes
shadowing appreciably smaller, especially at $x$ close to  $x= 10^{-2}$.

\bigskip
\begin{figure}[htbp]
\centerline{\epsfxsize=9.5cm \epsfbox{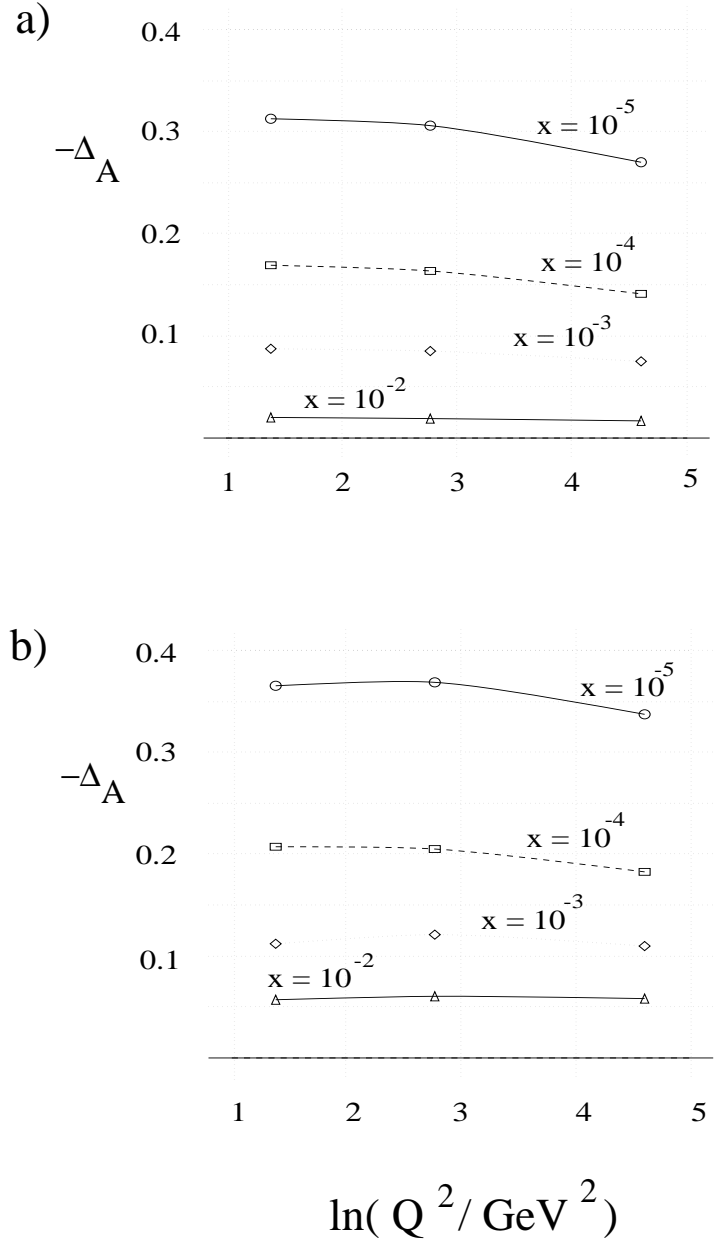}}
\bigskip
\caption{The ``triple pomeron'' contribution to $\Delta_A$ with
$\hat\alpha\not=0$ (a), and with $\hat\alpha=0$ (b), for Pb target.}
\end{figure}

\medskip
Several comments are in order:

(a) The rather weak $Q^2$ dependence of $\Delta_A$ seems puzzling if one
takes into account that in the dipole model the total (virtual) 
photon-hadron cross section decreases as $Q^{-1}$ at large $Q$
\cite{M94,NPR95,B96}. One would thus naively expect 
$\Delta_A \, \sim \, \langle f^2 \rangle / \langle f \rangle$ to
decrease also as $Q^{-1}$ which is another version of the
Bjorken-Gribov \cite{G70,B76} paradox. In the dipole model the paradox
is resolved (and approximate scaling of $\Delta_A$ recovered) due to the
cascade nature of the intermediate state which interacts inside the
nucleus. This cascade nature implies very strong correlations between
the dipoles in the onium representing the virtual photon so that the
double dipole density does not resemble at all the product of single
densities and both $\langle f^2 \rangle$ and $\langle f \rangle$
behave as  $Q^{-1}$ (apart from logarithmic corrections). The approximate
scaling of $\Delta_A$ follows. It is interesting to observe that this
explanation is radically different from that proposed in the parton model
(first paper of \cite{NZ91}, \cite{B76,FS89}).

(b) It should be pointed out that, although our calculation is based
on perturbative QCD (as is the whole dipole approach), it nevertheless
requires two ``non-perturbative'' parameters ($r_0$ and $n_{eff}$)
describing structure of the nucleon. These parameters were estimated
from the data on the total (virtual) photon-proton cross section as
they cannot be calculated in the dipole model. Particularly important
for the absolute value of the shadowing turns out to be $n_{eff}$, the
average number of ``onia'' forming the proton. This confirms the point
of view that the perturbative QCD alone cannot describe fully
deep inelastic scattering even at very small $x$ \cite{JDB96}. It is
rather remarkable, however, that the required ``non-perturbative''
input is so restricted.

(c) As was already pointed out in the main text, our calculation relies 
heavily on the calculation of the deep inelastic diffraction reported in
\cite{BP96a,BP96b}. This is a consequence of close relation between the
nuclear shadowing and diffraction (see
e.g. \cite{ABKSS96}). Nevertheless, it is worth to mention that --- due
to the power-law tail in the impact parameter dependence of the
quasielastic and triple pomeron amplitudes --- nuclear shadowing is
much more sensitive to the details of the effective cut-off in the
photon-nucleon impact parameter plane than is the total diffractive
cross section. At this point the two calculations are substantially
different.

(d) Our treatment of multiple scattering is only approximate. To
improve on this point, it is necesarry to calculate directly at least
the contribution of interaction with 3 nucleons which would give an
estimate of the accuracy of our approximation. This seems a feasible,
although probably a complicated calculation.

\begin{figure}[htbp]
\centerline{\epsfxsize=9.0cm \epsfbox{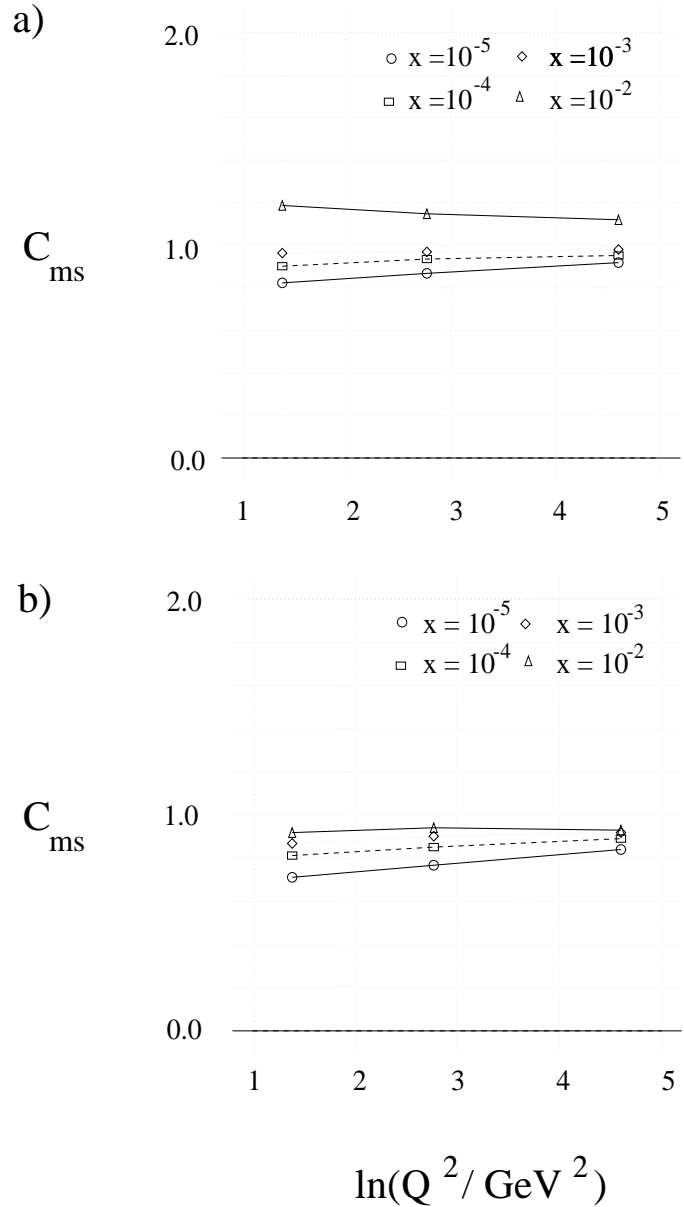}}
\bigskip
\caption{The correction factor $C_{ms}(x,Q^2)$ for higher than double
scatterings with $\hat\alpha\not=0$ (a), and with $\hat\alpha=0$ (b), for
Pb target.}
\end{figure}   

\acknowledgments
This work was supported in part by the KBN Grant No 2 P03B 083 08
(A.B. and W.C.), by the Stiftung fur Deutsch-Polnische
Zusammenarbeit project 1522/94/LN (W.F.), and by PECO grant from the
EEC Programme ``Human Capital Mobility'', Network ``Physics at High
Energy Colliders'' (Contract Nr: ERBICIPDCT 940613).  One of the
authors (W.C.) thanks Institute of Nuclear Physics in Krak\'ow for
financial support.

\vspace{2.0cm}

\appendix
\section{}

Here we give more details of some intermediate steps of calculations
of Section. 3.  First the integration over ${\bf \tilde b}$ and ${\bf
\tilde b'}$ in (\ref{eq10}). It can be done using the formula

\begin{equation}
\int\limits^\infty_{\hbox{\small max}(r,x)} \!\!\!d^2\tilde b
\, \tilde b^{-2} \,
\ln({\tilde b^2\over rx}) e^{-{\textstyle{1\over 2}} a_p\ln^2({\tilde b^2\over rx})} =
\pi\int\limits^\infty_0du(u+h)e^{-{\textstyle{1\over 2}} a_p(u+h)^2}={\pi\over a_p}
e^{-{\textstyle{1\over 2}} a_p h^2} \,\,,
\label{eqa1}
\end{equation}
where $h=\ln(r/x)$.

\medskip
The integration over $d^2rdz$ is done with the help of the
(asymptotic, $a_p\to 0$ ) formula 

\begin{equation}
\int\Phi(r)re^{-{\textstyle{1\over 2}} a_p\ln^2({r\over x})}d^2rdz=r_0
 e^{-{\textstyle{1\over 2}} a_p\ln^2({r_0\over x})}\,\,,
\label{eqa2}
\end{equation}
where $r_0=\int\Phi(r)rd^2rdz$. The argument justifying (\ref{eqa2})
goes as follows. We start with the identity

\begin{equation}
\left({r\over x}\right)\,e^{-{\textstyle{1\over 2}} a_p
\ln^2({r\over x})} = 
2\sqrt{\pi\over 2a_p}x^{\Delta_p}_p \int{d\gamma\over2\pi i}
\left({r\over x}\right)^\gamma
e^{-{\alpha N_c\over\pi}\chi(\gamma)\ln(x_p)}\,\,,
\label{eqa3}
\end{equation}
and write
\begin{equation}
\int\Phi(r)re^{-{\textstyle{1\over 2}} a_p\ln^2({r\over x})}d^2rdz
= 2\sqrt{\pi\over2a_p} x^{\Delta_p}_px 
\int{d\gamma\over2\pi i}e^{-{\alpha N_c\over\pi}\chi(\gamma)
\ln(x_p)} \int\Phi(r)\left({r\over x}\right)^\gamma d^2rdz\,\,\,.
\label{eqa4}
\end{equation}

Let us denote
\begin{equation}
\int\Phi(r)r^\gamma d^2rdz\equiv \tilde r^\gamma_0\,\,\,.
\label{eqa5}
\end{equation}
Of course $\tilde r_0$ depends on $\gamma$.  But, in the limit $a_p\to
0$, the main contribution to the integral (\ref{eqa4}) comes from the region
$\gamma\approx 1$ and we approximate $\tilde r_0$ by $r_0=\tilde
r_0(\gamma=1)$.  Using this we obtain from (\ref{eqa4})

\begin{equation}
\int\Phi(r) r e^{-{\textstyle{1\over 2}} a_p\ln^2({r\over x})}d^2rdz = 
2\sqrt{\pi\over2a_p}x^{\Delta_p}x\int{d\gamma\over2\pi i}
e^{-{\alpha N_c\over\pi}\chi(\gamma)\ln(x_p)}({r_0\over x})^\gamma\,\,.
\label{eqa6}
\end{equation}
We use again (\ref{eqa3}) and obtain (\ref{eqa2}).

\medskip   
Now the formula (\ref{eq14}):  Introducing (\ref{eqa1}) and 
(\ref{eqa2}) into (\ref{eq10}) we obtain

\begin{eqnarray}
\int d^2\tilde b d^2\tilde b'(\Sigma'_nP_nf_n(\tilde{\bf b})
f_n(\tilde{\bf b'})) &=&
{16\pi\alpha^4_{eff}\alpha N_c r^2_0\over x^{2\Delta_p}_p}
\left({a_p\over\pi}\right)
\int\limits^{c+i\infty}_{c-i\infty}{d\gamma\over2\pi i}\rho^{2-\gamma}
\beta^{-\alpha N_c\chi(\gamma)/\pi} \nonumber \\
& & \times\int dx_{12}dx_{02}
W(x_{12},x_{02})e^{-{\textstyle{1\over 2}} a_p(h^2_0+h^2_1)}\,\,,
\label{eqa7}
\end{eqnarray}
where $h_0=\ln({r_0\over x_{02}})$, $h_1=\ln({r_0\over x_{12}})$ .

\medskip
Next point is the integration over $dx_{02}dx_{12}$.  This we do as in
\cite{BP96a}  (although the accuracy of this procedure may be not very good).
Let us repeat the steps.  Since the integrand is symmetric in $x_{02} , 
x_{12}$ we have

\begin{equation}
I\equiv \int\limits^\infty_0dx_{12}dx_{02}W(x_{12},x_{02})
e^{-{\textstyle{1\over 2}} a_p(h^2_0+h^2_1)}
=2\int\limits^\infty_0dx_{12}e^{-{\textstyle{1\over 2}} a_ph^2_1}x^{\gamma-1}_{12}
\int\limits^{x_{12}}_0{dx_{02}\over x_{12}}e^{-{\textstyle{1\over 2}} a_ph^2_0}
F({x_{02}\over x_{12}})\,\,.
\label{eqa8}
\end{equation}
The second integral can be approximated (in the limit $a_p\approx 0$)
by $e^{-{\textstyle{1\over 2}} a_ph^2_1}V(\gamma)$, where $V(\gamma)$ is given below
Eq.(\ref{eq14}). Substituting this into (\ref{eqa8}) we obtain

\begin{equation}
I=2V(\gamma)\int\limits^\infty_0{dx_{12}\over x_{12}}x_{12}^\gamma
e^{-a_p\ln^2({r_0\over x_{12}})} \,\,.
\label{eqa9}
\end{equation}
This is an easy integral (gaussian, after substituting
$u=\ln({r_0\over x_{12}})$) and we obtain finally
\begin{equation}
I=2V(\gamma)r^\gamma_0\sqrt{\pi\over a_p}e^{{\gamma^2\over4a_p}}\,\,.
\label{eqa10}
\end{equation}
This completes the integrations in (\ref{eqa7}) and we obtain Eq. 
(\ref{eq14}).

\section{}

We comment here on our numerical calculations.  The contour integrals in
Eqs. (\ref{eq1}) and (\ref{eq14}) and the integral of the hypergeometric 
function defining $V(\gamma)$ were calculated with {\sl Mathematica}.

\medskip
For a fixed value of Bjorken`s $x$, the contour integral in (\ref{eq14})
becomes a function of two variables: $\rho$ and $\beta$.  Such a
two-dimensional function was tabulated (for $x=10^{-5}$,$x=10^{-4}$,
$x=10^{-3}$ and $x=10^{-2}$) and used as an input in the Fortran program
calculating the triple pomeron contribution (\ref{eq23}). Similarly, the
integral over $\hat z$ of the probability distribution
$\tilde\Phi(\hat z,\rho,Q)$ was first tabulated with the help of
{\sl Mathematica}, and then used as an input in the Fortran program
calculating (\ref{eq23}).

\medskip
{\sl Mathematica} was used once again to calculate and tabulate the 
integrals over $\rho$ of the Bessel functions in Eq. (\ref{eq26}).

\end{document}